\documentclass[aps,prl,twocolumn,showpacs,superscriptaddress,groupedaddress,nofootinbib]{revtex4-1} 
\usepackage{graphicx}
\usepackage{amsfonts,amssymb}
\usepackage{amsmath}

\begin{document}

\title{Opinion groups formation and dynamics : structures that last from non lasting entities}

\author{S\'ebastian Grauwin (1) and Pablo Jensen (2)} 
\address{1: Universit\'e de Lyon; IXXI, Rh\^one Alpes Institute of Complex Systems, 69364 Lyon; LIP, UMR CNRS 5668, INRIA; and ENS de Lyon, 69364 Lyon, France\\ 2: Universit\'e de Lyon; IXXI, Rh\^one Alpes Institute of Complex Systems, 69364 Lyon; and Laboratoire de Physique, UMR CNRS 5672, ENS de Lyon, 69364 Lyon, France}

\maketitle

We extend simple opinion models to obtain stable but continuously evolving communities. Our scope is to meet a challenge raised by sociologists of generating ``structures that last from non lasting entities''. We achieve this by introducing two kinds of noise on a standard opinion model. First, agents may interact with other agents even if their opinion difference is large. Second, agents randomly change their opinion at a constant rate. We show that for a large range of control parameters, our model yields stable and fluctuating polarized states, where the composition and mean opinion of the emerging groups is fluctuating over time.


\vspace{.5cm}
\section{Introduction}
\vspace{.2cm}

Several hundred papers have been published these last years by physicists on the dynamics of ``opinion'' group formation (for reviews, see Refs. \cite{castellano,lorenz}). While the relation to real human opinions is at most analogical, these simple models allow physicists to investigate a classical statistical physics topic : the formation of macroscopic states (here, of agents sharing similar opinions) from microscopic agents, whose opinions are initially randomly distributed. The main motivation of these studies was made explicit by Axelrod \cite{axelrod97}. If we assume that similar agents tend to become more similar by interacting, how comes that the real world shows an enduring diversity of groups, instead of convergence to a single opinion (``consensus'' state)? Depending on the topology and precise opinion imitation mechanisms and parameter values, these models yield different macroscopic states at equilibrium : fragmented (a state in which the opinions of the agents are uncorrelated), consensus and, more interesting, polarized. Obtaining a stable polarized state, where a finite number of macroscopic groups are formed, is the real scope of such models \cite{deffuant,krause}.

However, several studies have shown that the polarized structures obtained in these models are not really stable \cite{lorenz,klemm,mas,pineda,carletti2008}. Adding noise to the bounded imitation process, by allowing agents with very different opinions to interact with a non-zero probability, leads to convergence towards a consensus state\cite{lorenz,klemm,mas}. Two apparent exceptions have in fact added a mechanism specifically tailored to obtain stability. First, Kozma and Barrat \cite{barrat} showed that an adaptive network (where the social links of the agents are dynamically updated) is more stable than a static one, but they only tested its stability vis-\`a-vis an {\it asymmetric} noise where agent-agent links cannot be broken if agents' opinions are close enough. Here, we show that using a symmetric and more natural definition of noise, agents converge to consensus, even for an adaptive network. Second, M\"as et al \cite{mas} have recently introduced a different kind of noise (see the discussion below) as an ``individualization mechanism'' which, for a range of parameters, leads to stable polarized states. However, this noise seems an ad-hoc assumption, since it is specifically designed to break down the consensus cluster and avoid convergence.

In this paper, we review the stability of polarized states in several opinion models and prove that noise leads to consensus. We then introduce a new opinion model, which includes a turnover mechanism on agents' opinions and leads to polarized states which are robust. Our new model focuses on {\it dynamical} clusters, instead of looking for frozen polarized states \cite{barrat,holme}.

\vspace{.5cm}
\section{Model}
\vspace{.2cm}

\subsection{Deffuant bounded confidence model}
We build in this paper on Bounded Confidence opinion models \cite{deffuant,krause}. In these models, agents are characterized by their opinions, represented by a real in the $[0,1]$ interval and a set of links to other agents (for simplicity, we assume that agents are all linked, i.e. the social network is a complete graph). Agents can only interact with agents whose opinions are close to theirs, the range being given by a confidence threshold $d$.

In Deffuant original model, the dynamic rule is the following : at each elementary step $t$, two agents $i$ and $j$ are picked at random and they interact if their opinion difference $\Delta o = |o_i-o_j|$ is less than $d$. The interaction results in an update of their opinion, following

\begin{equation}
\begin{array}{lcl}
o_i(t+1) &=& o_i(t) + \mu(o_j(t)-o_i(t))\\
o_j(t+1) &=& o_j(t) + \mu(o_i(t)-o_j(t))
\end{array}
\end{equation}

where $\mu$ is a convergence parameter taken between $0$ and $0.5$. In this paper, we assume that $\mu=1/2$, meaning that interaction leads to convergence of the two agents to their average opinion.

\begin{figure}[h!]
\begin{center}
\includegraphics[width=0.45\textwidth]{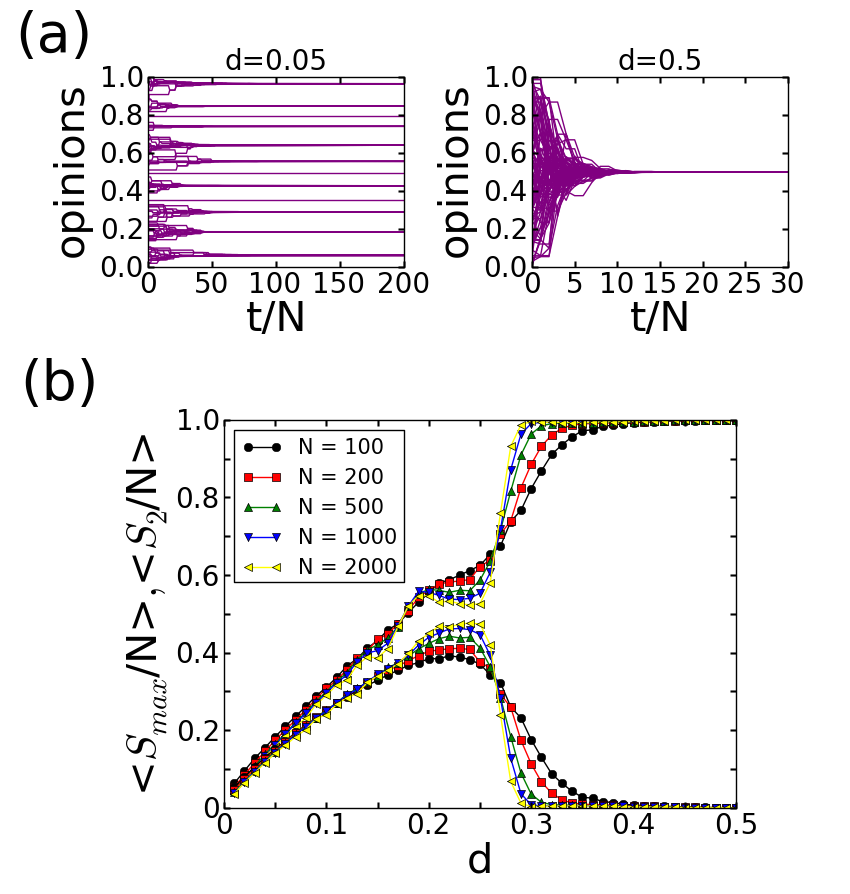}
\end{center}
\caption{\footnotesize{(Color online) {\bf Typical results observed in Deffuant's opinion model. (a)} Evolution of the agents' opinions, denoted by lines, for a system of $N=100$ agents with tolerance $d$ on a complete graph. In the initial state, agents' opinions are randomly distributed. The system converges towards a frozen state showing groups in which agents have all the same opinion. {\bf (b)} Relative size of the largest and second largest groups in the final state as a function of the tolerance threshold $d$ and for different system sizes.}}
\label{fig-deffuant}
\end{figure}

The main results of this model are well described in the literature \cite{castellano,deffuant} and summarized in Figs. \ref{fig-deffuant}. The dynamics leads to the formation of one or several groups of agents sharing the same opinion (Fig. \ref{fig-deffuant}a). The number and relative size of the groups depend on the parameter $d$. Final states are characterized by the relative size of the largest and the second largest groups (resp $<S_{max}/N>$ and $<S_2/N>$, see Fig. \ref{fig-deffuant}b).  

\subsection{Interaction noise}
Deffuant's model final states are crucially dependent on the sharp condition for interaction : $\Delta o < d$, which warrants that convergence will stop at some time, leading to stable polarized states. However, it is clear that such a sharp condition is unrealistic. It has already been shown \cite{lorenz,mas,bennaim} that the polarized state is not robust when perturbed e.g. by interaction noise or agents' heterogeneity. Here, we explore the stability of the results when the interaction rule is not sharp. We soften Deffuant's condition by introducing a natural, thermal noise which allows for a small probability of interaction between agents with opinion difference larger than $d$. Specifically, we define the probability of convergence $p_{conv}$ for two agents $i$ and $j$ with an opinion difference $\Delta o = |o_i - o_j|$ as: 

\begin{equation}
p_{conv} = \left[1 + \exp{\left(\frac{\Delta o/d - 1}{T}\right)}\right]^{-1}
\label{pconv}
\end{equation}

where parameter $T$ resembles a temperature and characterizes the steepness of the convergence. A small $T$ value means that the transition is steep, while a large value indicates that an opinion difference is not important. As we will show, the introduction of $p_{conv} > 0$ for $\Delta o > d$ leads to convergence as the only final state. We argue that this form (Eq. \ref{pconv}) of noise is more natural than that introduced by Kozma and Barrat \cite{barrat} in their adaptive network model. The point is that their noise is similar to ours {\it only for} $\Delta o > d$ (in their model, noise plays no role when agents are close enough in opinion space).

\subsection{Opinion noise as turnover}
The second ingredient of our model leads to a never ending dynamics. Specifically, at a given rate $\nu$, we remove an agent from the system and replace it with a new one endowed with a new, {\it random} opinion. The number of agents in the system remains constant. This ingredient, very similar to the noise introduced in Axelrod's model by \cite{klemm}, or in Deffuant's model by \cite{pineda2009,nyczka2011}, can be interpreted as a `death' of the agent and a `birth' of a new one, or as an opinion noise. It is much simpler and natural than the noise introduced by \cite{mas}, which depends on the size of the group the agent belongs to.

\subsection{Model summary}

To summarize, on each elementary step $t$ we do the following:
\begin{itemize}
\item 1. Pick an agent $i$ at random. With probability $\nu$, update agent $i$: 
$o_i$ takes a new random value between $0$ and $1$.
\item 2. Pick two agents $j$ and $k$ at random. With probability 
$$p_{conv} = \left[1 + \exp{\left(\frac{\Delta o/d - 1}{T}\right)}\right]^{-1}\,$$ 
where $\Delta o = |o_j(t)-o_k(t)|$, the opinions of $j$ and $k$ converge to their average according to 
\begin{equation*}
\begin{array}{lcl}
o_i(t+1) &=& o_i(t) + (o_j(t)-o_i(t))/2\\
o_j(t+1) &=& o_j(t) + (o_i(t)-o_j(t))/2
\end{array}
\end{equation*}
\end{itemize}

Our model is based on four parameters: two which are common to most bounded-confidence opinion models (number $N$ of agents, tolerance threshold $d$) and two which are specific to our model (interaction noise $T$, opinion noise $\nu$). 
In the following, we will refer to a set of $N$ successive elementary steps as an {\it iteration}. This normalization of the time scale is obviously more adapted to follow the dynamic evolution of the agents. Indeed, on average each agent is picked for a tentative update during an iteration. It follows for example that the agents' lifetime expectancy ( $\sum_{n=0}^{\infty}n\nu\left( 1-\nu \right)^{n} = \frac{1}{\nu}-1$ iterations) is independent from the number $N$ of agents.

\subsection{Defining groups}

Defining groups in Deffuant's model final states is straightforward since groups are cliques of agents sharing the same opinion. Here, because of the introduction of opinion noise (when $\nu > 0$), the distribution of opinions is more diffuse than in Deffuant's case and groups are less clear cut. There are several ways to define an opinion group in a more general context. We choose here to follow the definition of proposed by \cite{barrat}, based on the notion of {\em communicating agents}. Two agents $i$ and $j$ are said to be communicating agents if their opinion difference is within the tolerance value ($\Delta o = |o_i-o_j| < d$). An opinion group is then defined as a set of agents all linked to each other through a path of communicating agents. This corresponds to the notion of a channel of communication through which ideas can be exchanged between the agents. 

\vspace{.5cm}
\section{Results for limit cases}
\vspace{.2cm}
In this section, we study separately the effects of the introduction of interaction noise and turnover in the standard Deffuant model.

\subsection*{Effect of interaction noise}
We start by assuming that there is no agent turnover ($\nu=0$). What is the effect of introducing noise in the interaction process ($T>0$)?

\begin{figure}[h!]
\begin{center}
\includegraphics[width=0.45\textwidth]{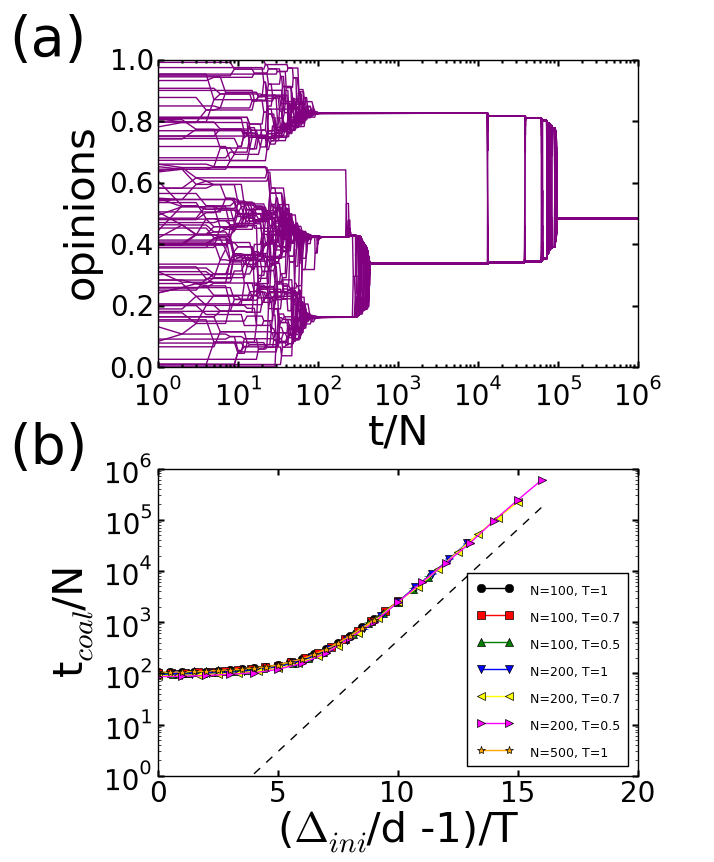}
\end{center}
\caption{\footnotesize{{(Color online) \bf Interaction noise pushes the system towards consensus. (a)} Evolution of agents' opinions towards consensus in case of interaction noise ($N=100$, $d=0.06$, $T=0.5$ and $\nu=0$). {\bf (b)} Typical convergence time to consensus $t_{coal}$, starting with two artificially built groups with opinions $0$ and $\Delta_{ini}$. Each point corresponds to the average over $100$ simulations. The dashed line shows the slope of the exponential function ($t_{coal}/N=\exp x$, where $x = [\Delta_{ini}/d-1]/T$).}}
\label{fig-interaction-noise}
\end{figure}

Fig. \ref{fig-interaction-noise}a shows an example of the observed dynamics. In a first step, there is local convergence of agents with similar opinions (at the $d$ scale), as in the standard Deffuant model. On a longer time-scale, these groups interact through ``interaction leaks'', i.e. pairs of agents than manage to converge thanks to the noise (since $p_{conv}$ is strictly positive even for $\Delta o > d$). As groups get closer by partial convergences, there is an acceleration of the convergence since it gets exponentially easier. In the long run, the system reaches a consensus state.

It is rather straightforward to understand that a Deffuant model with interaction noise always leads to consensus, whether the social network is static (which is the case here) - provided the network is connected, or whether the network is dynamic (which is the case in Barrat's model \cite{barrat}) - provided the probability for an agent to break a link and to rewire it to a given agent in the network is non zero. We propose in the following an argument supporting this claim.

\begin{flushleft} \hspace{0.5cm} \underline{Argument}\\\end{flushleft}
\vspace{-2mm}
Define $o_{max} = max\{o_i\}$ and $o_{min} = min\{o_i\}$ as representing the extremal opinions among agents at a given time. In the absence of agent turnover ($\nu = 0$), $|o_{max}(t)-o_{min}(t)|$ is a decreasing positive function which converges towards a lower limit, which is necessarily $0$. Indeed, in the presence of interaction noise ($T>0$):
\begin{itemize}
\item {\em for a static, connected network}, while $o_{max} > o_{min}$, any agent with opinion $o_{max}$ (resp $o_{min}$) can only decrease (resp increase) his opinion by interacting with another agent with a lower (resp larger) opinion. 
In the case of a complete network, the probability to do so at a given iteration is always strictly positive since an agent can interact with any other (e.g., an agent with opinion $o_{max}$ can interact with an agent of opinion $o_{min}$), which ensures that it does happen in the long run. 
In a more general case, it may happen that a given iteration an agent e.g. with opinion $o_{max}$ only have neighbours with opinion $o_{max}$. But since the network is connected, at least one agent with opinion $o_{max}$ must have at least one neighbour with a lower opinion. This condition ensures that all the agents with opinion $o_{max}$ (resp $o_{min}$) interact with an agent with a lower (resp larger) opinion in the long run.  
\item {\em for a dynamic network} where the agents are able to break their links and rewire them at random (which is the case in \cite{barrat}),
the same reasoning holds. While $o_{max} > o_{min}$, an agent with opinion $o_{max}$ either has a neighbour with a lower opinion or can break one of his links to rewire it to an agent with a lower opinion. Provided the probability of this last process is non-zero (which is NOT the case in \cite{barrat}), an agent with opinion $o_{max}$ (resp $o_{min}$) thus always has a non-zero probability to interact with an agent with a lower (resp larger) opinion, which ensures that it does happen in the long run.$\square$  
\end{itemize}

In the dynamic network model presented by \cite{barrat}, agents are unable to break bonds between neighbors with close opinions ($\Delta o < d$), which leads to the stability of the polarized state. Indeed, once several groups (corresponding to the connected parts of the network) are formed, intra-group bonds can never be broken and therefore the groups can never connect to coalesce. Introducing our more symmetric definition of noise in the link-breaking probability of their model would lead to a single, consensus cluster, as the equilibrium state.

Can we compute the characteristic coalescence time? Fig. \ref{fig-interaction-noise}a suggests that the time needed for two groups to coalesce increases very rapidly when opinion distance increases. To quantify this intuition, let us define $t_{coal}(N,d,\Delta_{ini},T)$ as the average number of elementary steps needed to reach consensus or {coalescence time}, starting with two groups of $N/2$ agents, one with opinion $0$ and another with opinion $\Delta_{ini}>0$. Since the interaction rule depends on $\Delta o / d = \Delta (o/d)$ (see Eq. \ref{pconv}), agents' opinion can be normalized by $d$ and the dependence of the coalescence time becomes $t_{coal}=t_{coal}(N,\Delta_{ini}/d,T)$. To investigate further the dependence of $t_{coal}$ with the model parameters, we performed a series of $100$ simulations for different values of $(N,\Delta_{ini}/d,T)$, measuring the average time needed to reach consensus \footnote{Consensus corresponds to all the agents sharing the same opinion $o_{\infty} \equiv 1/N\sum_i o_i$ (in case of no turnover, the mean opinion is constant in time). Note that if one defines $l(t)=\sum_i|o_i(t)-o_{\infty}|$ for any given time $t$, analytically speaking $\lim\limits_{t \to \infty}(l(t))=0$ but $l(t)>0$ for any finite $t$ as long as $N>2$. In practice, the measured freezing time corresponds to $t^*$ such that $l(t^*) < \epsilon$, where $\epsilon$ is the precision of the computer float representation (in our case, $10^{-16}$).}. 
Simulation results are displayed on Fig. \ref{fig-interaction-noise}b and fall on a single curve:
\begin{equation}
t_{coal}/N = f([\Delta_{ini}/d-1]/T)
\end{equation}

First, note that the linear dependence of $t_{coal}$ with $N$ arises from the proportionality of the number of elementary steps required to pick all agents. Second, we address the two regimes shown by Fig. \ref{fig-interaction-noise}b for different values of $x\equiv [\Delta_{ini}/d-1]/T$. 
For small values of $x$ (roughly $x < 7$), simulations show $t_{coal}/N \sim 100$, which corresponds roughly to the time needed for initially random opinions to converge locally to a single cluster (see Fig. \ref{fig-deffuant}a). In the case where $T \ll 1$, this regime corresponds to $\Delta_{ini} < d$ i.e. a quasi-Deffuant regime in which all the agents are initially within each other tolerance threshold. For large values of $x$ (roughly $x > 7$), simulations show that $t_{coal}/N$ depends exponentially on $x$. In the case where $T \gg 1$, this regime corresponds to $\Delta_{ini} > d$ i.e. a quasi-Deffuant regime in which the two groups of agents are initially not within each other tolerance threshold. In that case, since the interaction probability increases rapidly (exponentially) as the opinion difference between agents decreases, the limiting time for reaching consensus is a normalized factor of the expected time needed for the first interaction between two agents belonging to the two groups, i.e. with an opinion difference $\Delta_{ini}$: 
\begin{eqnarray}
t_{first} &=& \sum_{t=0}^{\infty} t p_{conv}(x) \left(1 - p_{conv}(x)\right)^{t} \nonumber\\  
 &=& \frac{1}{p_{conv}(x)}-1  \nonumber\\
 &=&\exp{x},
\end{eqnarray} 

In summary, there are two regimes, the first one $t_{coal}/N \sim constant$ dominated by the minimum time needed by a set of agents all interacting with each other to reach consensus, the other $t_{coal}/N \sim \exp([\Delta_{ini}/d-1]/T)$ dominated by the time needed for the first interaction between two agents with a large opinion difference to occur.

\subsection*{Effect of turnover}
We now switch off the interaction noise ($T=0$) and study the effect of turnover ($\nu>0$).

\begin{widetext}

\begin{figure}[h!]
\begin{center}
\includegraphics[width=0.9\textwidth]{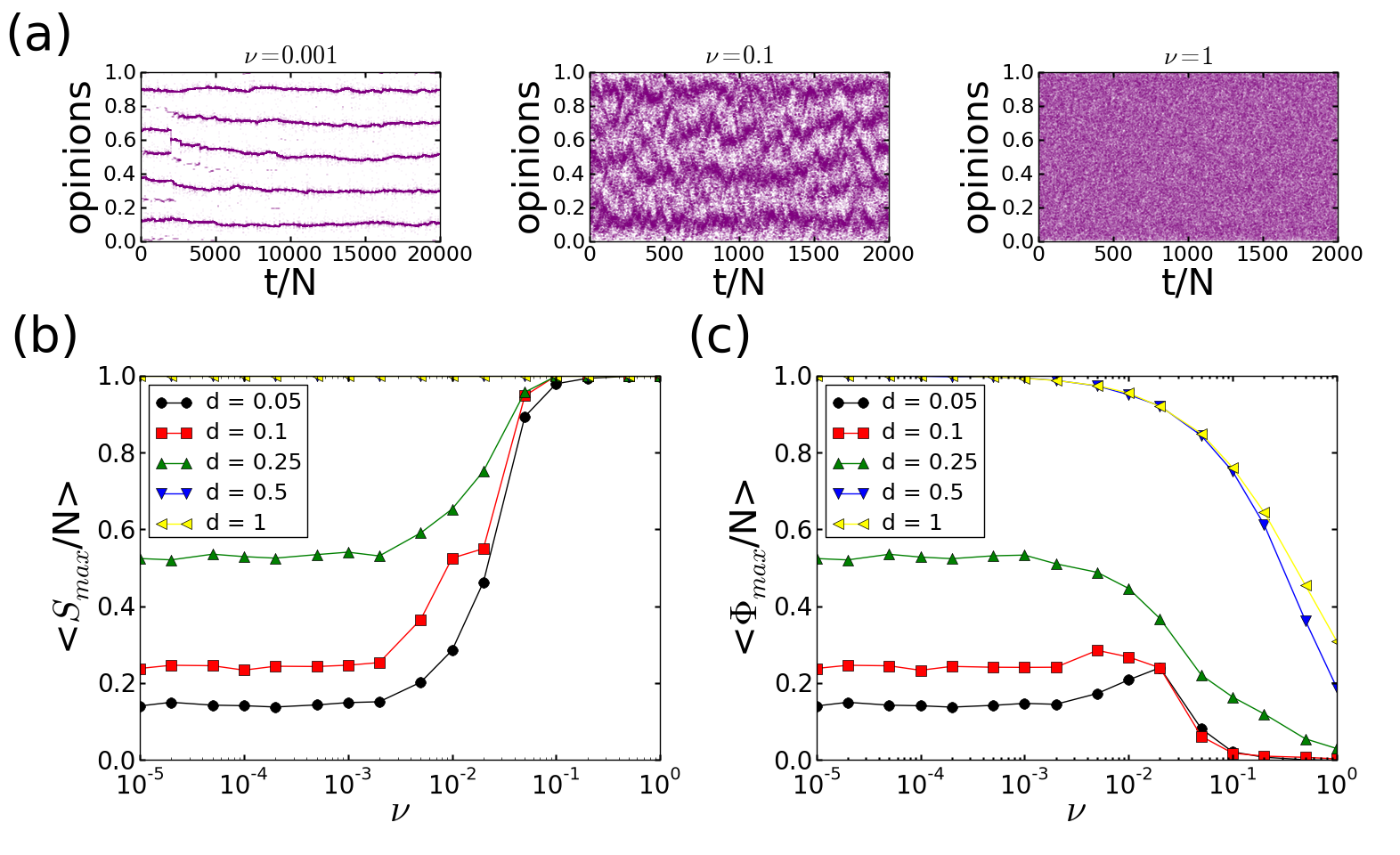}
\end{center}
\caption{\footnotesize{(Color online) {\bf Order / disorder transition induced by turnover. (a)} Evolution of agents' opinions for different values of turnover ($N=200$, $d=0.1$, $T=0$). {\bf (b)} Relative size of the largest opinion cluster. {\bf (c)} Values of the order parameter $\Phi_{max}/N$ as a function of $\nu$. The values of $S_{max}/N$ and $\Phi_{max}/N$ are averaged over a sample of $1000$ snapshots spread over at least $10$ agent's lifetimes, the averaging process beginning after $10$ lifetimes ($N=200$, $T=0$).}}
\label{fig-opinion-noise}
\end{figure}

\end{widetext}

Fig. \ref{fig-opinion-noise}a presents three examples of the dynamics obtained for $d=0.1$ and different values of turnover. When $\nu = 1$, agents change opinion at every iteration and, unsurprisingly, no collective structure emerges. Instead, we observe a homogeneous distribution of opinions. For low turnover values ($\nu = 10^{-3}$), the opinions of the agents are squeezed on a few values as in the usual Deffuant model. Groups are at opinion distances close to $\sim 2d$ as in the standard ($\nu =0$) case, with some apparently randomly distributed agents in between. The number of groups is rather stable and the mean opinion of a group fluctuates. Groups seem to move in a random walk in both the number of agents and average opinion, at least for short times. The intermediate case ($\nu = 0.1$) shows some structure but the overall picture is rather noisy.

Can we characterize the order-disorder transition with an order parameter? Fig. \ref{fig-opinion-noise}b shows the relative size of largest cluster $<S_{max}/N>$ (i.e., the usual order parameter, see Fig. \ref{fig-deffuant}b) as a function of $\nu$. This parameter does well for low values of $\nu$ but it turns out to be maximal for $\nu=1$ since, when the opinions are (randomly) spread over all the [0,1] interval, there exists a communication path among any pair of agents and $<S_{max}/N>=1$. Therefore, $<S_{max}/N>=1$ is not a good {\em order} parameter for this transition, as it does not take into account the diversity of opinions inside a given group.

To account for the intrinsic order of a group $g$, we introduce 
\begin{equation}
\psi_g \,=\, 1 - 3<\Delta o>_g
\end{equation}
where $<\Delta o>_g = \frac{2}{S_g(S_g-1)}\sum_{i\in g,\,j \in g}|o_i-o_j|$ is the mean difference of opinion between two agents of group $g$. This definition of $\psi_g$ ensures that $\psi_g=1$ for a coherent group (without opinion dispersion) and $\psi_g=0$ for a group of agents whose opinions are randomly distributed between $0$ and $1$.

We then introduce $\Phi_{max}$ to combine the two informations: the organization of agents in groups ($S_{max}$) and the internal structure of the largest group ($\psi_{max}$). This leads to the order parameter:
\begin{eqnarray}
\Phi_{max} &=& S_{max}\psi_{max}
\end{eqnarray}

Fig. \ref{fig-opinion-noise}c shows that $\Phi_{max}$ is indeed a good order parameter to quantify the order-disorder transition with $\nu$. 

The transitions observed in Figs. \ref{fig-opinion-noise}b and \ref{fig-opinion-noise}c can be understood as a competition between the update process and the imitation process. Indeed, in the limit $\nu \to 0$, the update process has almost no influence. Groups roughly corresponding to those generated by Deffuant's model are obtained. A small turnover does not impact the organization into groups where the agents all share the same opinion, but it induces however some fluctuations of the groups' mean opinion on large temporal scale (see Fig. \ref{fig-opinion-noise}a).
For median values of $\nu$, the update process produces agents with intermediate opinions which create communication channels between the original groups (see the case $\nu = 0.1$ in Fig. \ref{fig-opinion-noise}a). While the opinions of the agents can still be locally concentrated, groups are formed by two or more of the original groups plus the agents with intermediate opinion linking them. The size of the largest group $S_{max}$ increases with $\nu$ and so does the dispersion of opinions within these groups. The value of $\Phi_{max}$ reflects the combination of these two effects (see  Figs. \ref{fig-opinion-noise}b and \ref{fig-opinion-noise}c).   
In the limit $\nu = 1$, the turnover process generates a lot of opinion dispersion, ensuring the existence of communication channels between each pair of agents. A single group of maximum size $S_{max} = N$ is detected. Note that $\Phi_{max}$ can be strictly positive, i.e. the distribution of opinions is not completely random as one could have expected. Indeed, the imitation mechanism still ensures that at each iteration, a given proportion of agents interacts and converges in opinion space. This proportion obviously increases with Deffuant's threshold $d$, thus creating more local consensus between agents (see Fig. \ref{fig-opinion-noise}c). 

To summarize, two different transitions have to be distinguished:
\begin{itemize}
\item a communication transition due to the constant presence of agents with intermediate opinions linking different opinion groups. 
\item an order / disorder transition due to random opinion dispersion.  
\end{itemize}

The first type of transition occurs only for (roughly) $d<=0.25$ since at least two groups should exist in the limit $\nu \to 0$. Let us analyse the creation of a communication channel between a group of $2dN$ agents of opinion $o_1$ and a group of $2dN$ agents of opinion $o_2 = o_1 + 2d$. At a given elementary step, the agent picked by the dynamic process will be updated with an intermediate opinion $o\in [o_1,o_2]$ with a probability $2d\nu$. On the other hand, at a given elementary step, an agent with an intermediate opinion will be `reabsorbed' by interacting with a member of the closest group with a probability $ 2(2dN/N)(1/N) = 4d/N$. Comparing these two values, one can predict that the communicating transition can be associated with by a characteristic transition rate $\nu_d \sim 2/N$.

The second type of transition is similar to the usual competition between opinion dispersion (characteristic time $1/\nu$, i.e. an agent's lifetime) and imitation (characterized by the number $\tau_c \sim 100$, of iterations needed to reach consensus starting with randomly distributed opinions, see Fig \ref{fig-deffuant}). The order / disorder transition can thus be associated with a characteristic transition rate $\nu_c = 1/\tau_c$. 

\begin{figure}[h!]
\begin{center}
\includegraphics[width=0.45\textwidth]{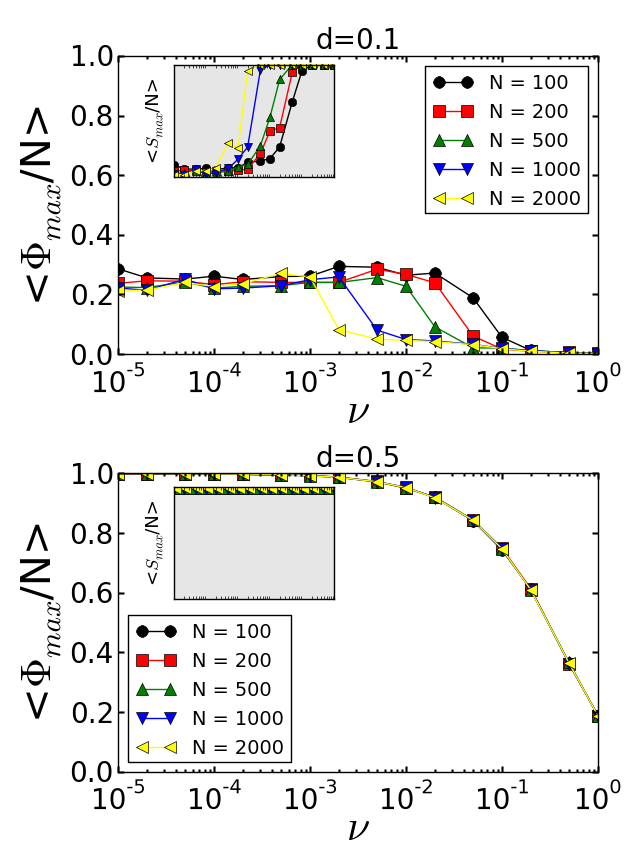}
\end{center}
\caption{\footnotesize{(Color online) {\bf Influence of the number $N$ of agents} on the order parameter $\Phi_{max}/N$ [Inset: $S_{max}/N$] for different values of $\nu$ and $d$. The values of $S_{max}$ and $\Phi_{max}/N$ are averaged over a sample of $1000$ snapshots spread over at least $10$ agent's lifetimes, the averaging process beginning after $10$ lifetimes ($T=0$).}}
\label{fig-size-effects}
\end{figure}

These two kinds of transition are shown in Fig. \ref{fig-size-effects}. Indeed, the case $d=0.5$ shows a pure order / disorder transition where the curve $\Phi_{max}(\nu)$ is independent from the number $N$ of agents. On the contrary, the case $d=0.1$ displays communicating transitions, which is reflected by the dependence of the transition rate with $N$. Of course, in most cases, the order / disorder transition is induced by the two mechanisms acting together. In the following, we will use the single notation $\nu_0(N) = \min(\nu_c,\nu_d(N))$ to denote the characteristic transition rate.

\begin{widetext}
 
\begin{figure}[h!]
\begin{center}
\includegraphics[width=0.75\textwidth]{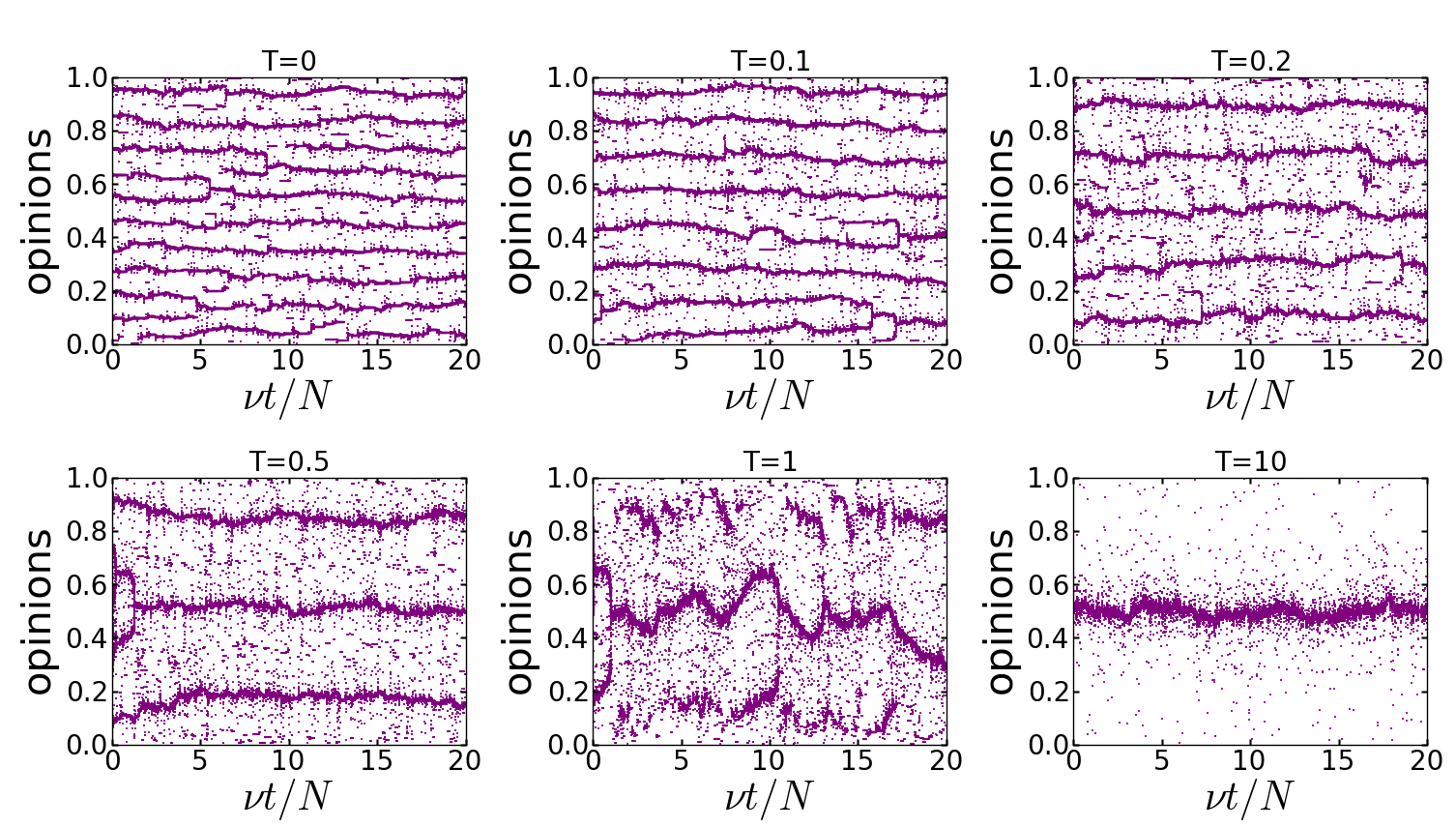}
\end{center}
\caption{\footnotesize{(Color online) {\bf Evolution of agents' opinions} for different levels of interaction noise, in the model including turnover. For a given range of the parameters, the coalescing force of the interaction noise can be stabilized by the dispersion of opinions generated by the turnover, so that the system always remain in a polarized state. The time axis has been normalized in typical agent's lifetime units.  ($N=100$, $d=0.05$, $\nu=0.001$) }}
\label{fig-TN}
\end{figure}

\end{widetext}

\vspace{.5cm}
\section{Results in the general case}
\vspace{.2cm}

Let us investigate the results obtained when the two opposing ingredients are combined, as interaction noise pushes towards consensus while turnover leads to opinion dispersion.

Fig. \ref{fig-TN} displays examples - with a turnover rate $\nu =10^{-3}$ and a range of interaction noise $T$ - showing that this combination can lead to a dynamical equilibrium where groups persist. While the turnover process induces some fluctuations in their number of agents and average opinion, these groups last on time scales much larger than a typical agent's lifetime. 

\begin{figure}[h!]
\begin{center}
\includegraphics[width=0.45\textwidth]{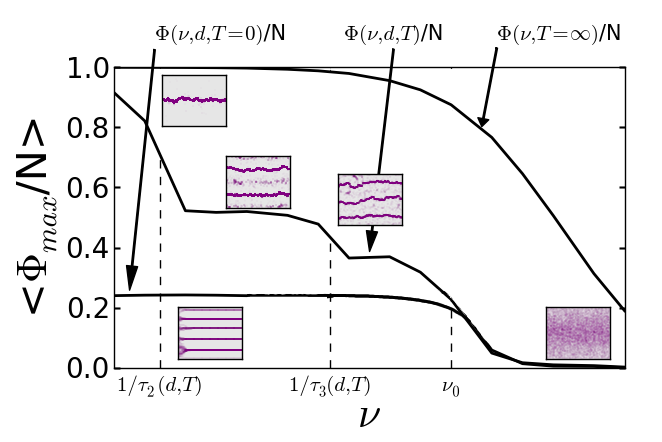}
\end{center}
\caption{\footnotesize{(Color online) {\bf Qualitative schema} showing the dependence of $<\Phi_{max}/N>$ with the different parameters. Refer to the main text for explanations.}}
\label{fig-interpr}
\end{figure}

\begin{widetext}

\begin{figure}[h!]
\begin{center}
\includegraphics[width=0.9\textwidth]{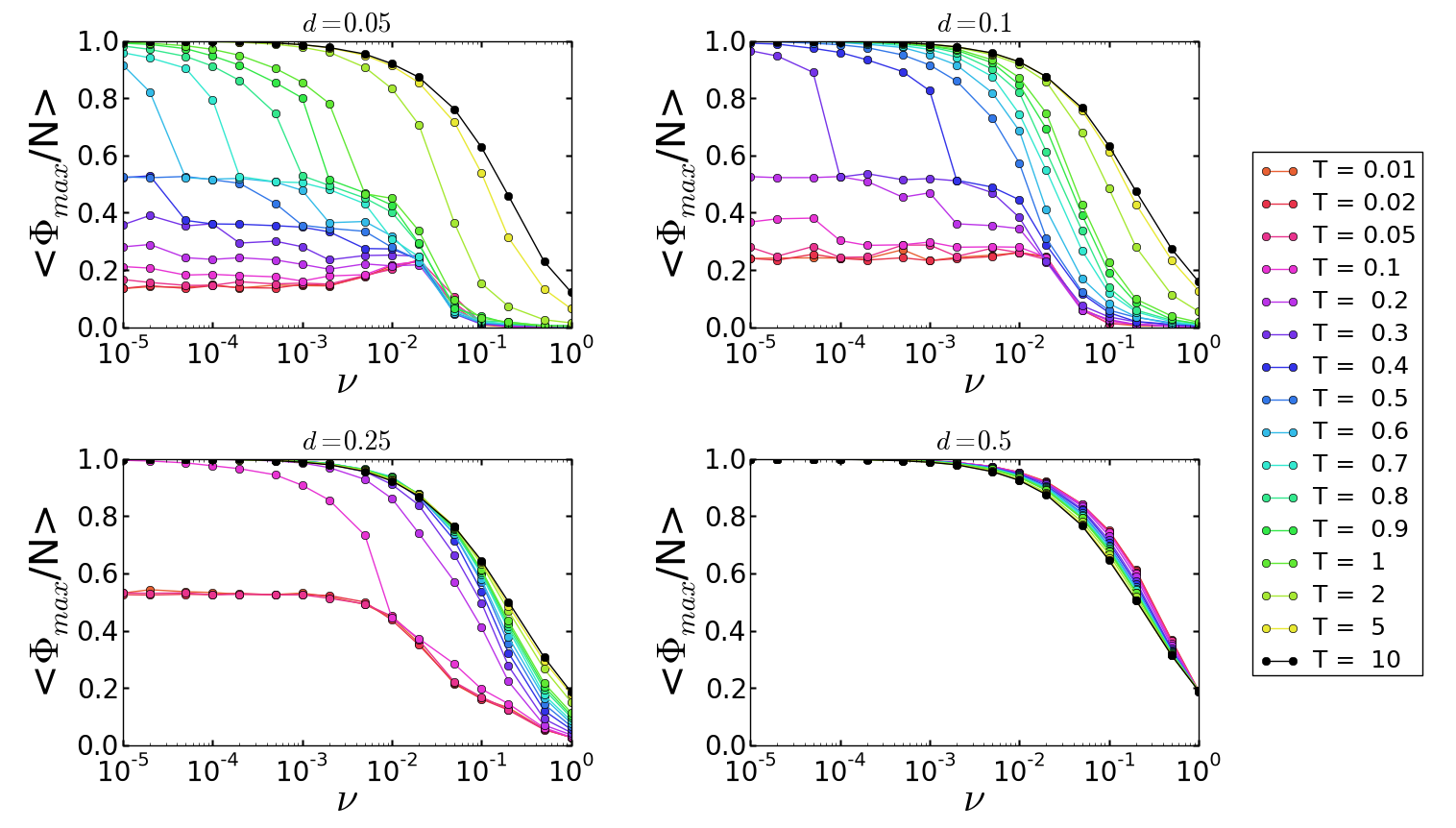}
\end{center}
\caption{\footnotesize{(Color online) {\bf Phase diagrams} Values of the order parameter $\Phi_{max}/N$ as a function of $\nu$ and $T$ for different threshold parameters $d$. The values of $\Phi_{max}/N$ are averaged over a sample of $1000$ snapshots spread over at least $10$ agent's lifetimes, the averaging process beginning after $10$ lifetimes ($N=200$). }}
\label{fig-phi}
\end{figure}

\end{widetext}

We now analyze in detail the phase diagram obtained for different values of the noise parameters $\nu$ and $T$. Fig. \ref{fig-interpr} shows a qualitative phase diagram summarizing the different regimes which are found in simulations and characterized in Fig. \ref{fig-phi}. 
The first limiting case is obtained for $T \gg 1$, i.e. when all pairs of agents interact with a probability $p_{conv} (\sim 1/2$), independent of their opinion difference $\Delta o$ and of the threshold parameter $d$. This situation is similar to the one obtained in the previous section in case $T=0$ and $d=1$ (see Fig. \ref{fig-opinion-noise}), for which all pairs of agents interact with a probability $p_{conv=1}$. It leads to the upper limit curve of Fig. \ref{fig-interpr}, showing a continuous transition from order to disorder with characteristic turnover rate $\nu_0$. As can be checked on Fig. \ref{fig-phi}, this limit curve is the same for the four values of $d$ chosen in the displayed examples.

The opposite limit case corresponds to $T=0$ (lower curve), which was presented in the previous section (see Fig. \ref{fig-opinion-noise}). The order / disorder transition takes place for $\nu \sim \nu_0$. For $\nu \ll \nu_0$, the dynamics leads to roughly $1/(2d)$ groups without opinion dispersion, leading to $\Phi_{max} \simeq 2d$. Hence, the representative curve depends on the value of the threshold parameter $d$.
 
The curves corresponding to intermediate cases (finite non-zero interaction noise $T$) displayed on Fig. \ref{fig-phi} are characterized by successive transitions from one plateau value of $\Phi_{max}$ to another, each of these plateaus corresponding to a given number of groups (for instance, $<\Phi_{max}/N> \sim 1/2$ when agents are gathered in two main groups). The origin of these transitions can be understood as successive equilibria between the dispersive force due to the turnover rate and the coalescing force due to the interaction noise, whose amplitude depends on the number $n$ of groups.

Indeed, let us define $\tau_{n}(d,T)$ as the expected number of iterations needed to reach a typical polarized state with $n-1$ groups from a typical polarized state with $n$ groups thanks to the imitation process only (i.e. without turnover)\footnote{by `typical', we imply a state where each group contains roughly the same number of agents and where the opinion difference between groups is characterized by a parameter $\Delta o(n)$}. For example, $\tau_2(d,T)$ corresponds to the definition of the time $t_{coal}/N([\Delta_{ini}/d-1]/T)$ needed for two groups to coalesce introduced section 3, but for a specific value of $\Delta_{ini}$. The outcome of the dynamics can be understood by comparing characteristic times:

\begin{itemize}
\item When $\nu \gg \nu_0$, the dispersive force dominates and there is not any group structure.   
\item When $\nu \ll \nu_0$ and $\tau_{n}(d,T)^{-1} \ll \nu \ll \tau_{n+1}(d,T)^{-1}$, there is an equilibrium between the two forces leading to a stable structure of $n$ opinion groups and  $<\Phi_{max}/N> \sim 1/n$.
\item When $\nu \ll \nu_0$ and $\nu \ll \tau_{2}(d,T)^{-1}$, the coalescing force dominates, consensus is obtained and $<\Phi_{max}/N> \sim 1$.
\end{itemize}

The last point raised here implies that for $T>0$, the limit of the order parameter $\Phi_{max}/N$ when $\nu \to 0$ is always $1$, as shown on the qualitative schema Fig. \ref{fig-interpr}. However, in the same way that $\tau_{2}(d,T)$ varies exponentially with $1/T$ (as shown previosly), one can expect an exponential variation of $\tau_{n}(d,T)$ with ($1/T$) for any $n \geq 2$. This explains why the curves displayed on Fig. \ref{fig-phi} do not always show all the transitions until $\Phi_{max}/N = 1$. To do so would have required to run simulations for much lower values of the turnover parameter $\nu$. 


\vspace{.5cm}
\section{Discussion and conclusion}
\label{discussion}
\vspace{.2cm}

In this paper, we have presented a new opinion model showing polarized states which are robust to noise in the interaction process. The groups are dynamic, since they are constantly renewing their members, and yet they keep an identity, represented here by the average opinion which constantly fluctuates.

Note that the interaction noise we have introduced (Eq. \ref{pconv}) is similar to a standard thermal noise and has not been specifically tailored to prevent consensus (it actually leads to consensus in the absence of agents' turnover). This is in contrast with M\"as et al. \cite{mas}, who also obtain dynamic groups by introducing a specifically designed random change of opinion whose amplitude depends on the size of the group the agent belongs to. This noise preferentially breaks big clusters, which can be interpreted as an ad-hoc mechanism to prevent consensus. Kozma \& Barrat \cite{barrat} have argued that an adaptive network, where links are continuously rewired, is more robust with respect to interaction noise, but as we have shown this is only true for the kind of non symmetric interaction noise they use. Pineda et al. \cite{pineda,pineda2009} have introduced an ``opinion diffusion'' which is similar to our turnover in the limit of a large ``diffusion length'', but these authors have not studied the influence of interaction noise on the dynamics of their model. Nyczka \cite{nyczka2011} carefully studied how turnover leads to spontaneous transitions between different numbers of clusters, but, again, without interaction noise. Finally, Carletti et al \cite{carletti2008} have introduced opinion noise and interpreted it as birth and death of agents, but in a model with a complicated interaction noise relying on ``affinity score''. Moreover, they focused their attention on the transition between a single opinion cluster or a fragmented phase.

It is tempting to draw an analogy with real social groups, which are also constantly evolving yet retain an identity, and sometimes last longer than the agents' lifetime. However, as the [0,1] real numbers used here bear little similarity to actual opinions and the imitation mechanisms are too simple, the analogy may be more misleading than informative.

We suggest that the {\em artificial} societies invented in these models way may be useful to test or improve the conceptual tools developed by social scientists to understand some aspects of real societies. Take for example one fundamental question in sociology, already raised by Georg Simmel in 1898 \cite{simmel} : how can ``The Persistence of Social Groups'' be explained? For him, the key factor is that the ``displacement of one generation by the following does not take place all at once. By virtue of this fact it comes about that a continuity is maintained''. This paper has created a simple artificial society which shows an analogous phenomenon, i.e. structures that last from non lasting entities because continuity is maintained in opinion space. It is now up to sociologists to say if analyzing this simple society leads them to a better understanding of real ones \ldots Finally, this model, which shows rich group dynamics, could also be used as a controlled group dynamics generator, to test e.g. algorithms for dynamic communities detection such as \cite{palla}.\\

We are happy to acknowledge discussions with Bruno Latour, Tommaso Venturini, Paul Girard and Dominique Boullier from MediaLab (Sciences Po, Paris), Eric Bertin and Abdellah Fourtassi (Laboratoire de Physique, ENS de Lyon) and Guillaume  Beslon (INSA Lyon).


\end{document}